\documentclass[aps,prd,amsmath,amssymb]{revtex4}

\usepackage{graphicx}
\usepackage{dcolumn}
\usepackage{bm}

\def\Journal#1#2#3#4{{#1} {\bf #2}, #3 (#4)}

\def\AM{{\em Acta Math.}}
\def\AP{{\em Ann. Phys.}}

\def\CMP{{\em Comm. Math. Phys.}}
\def\IJMP{{\em Int. J. of Mod. Phys.}}

\def\NPB{{\em Nucl. Phys.} B}

\def\PLB{{\em Phys. Lett.} B}

\def\PRA{{\em Phys. Rev.} A}
\def\PRB{{\em Phys. Rev.} B}
\def\PRD{{\em Phys. Rev.} D}

\def\PREPC{{\em Phys. Rep.} C}

\def\RMP{{\em Rev. Mod. Phys.}}

\newcommand{\be}{\begin{equation}}
\newcommand{\ee}{\end{equation}\noindent}
\newcommand{\eei}{\end{equation}}
\newcommand{\bea}{\begin{eqnarray}}
\newcommand{\eea}{\end{eqnarray}\noindent}
\newcommand{\eeai}{\end{eqnarray}}

\newcommand{\hf}{\frac12}
\newcommand{\nn}{\nonumber\\}
\def\eq#1{(\ref{#1})}
\def\la{\langle}
\def\ra{\rangle}
\def\cd#1{{\cal D}[#1]}
\def\tr{{\mathrm {Tr}}~}
\def\ord#1{{\cal O}\left(#1\right)}

\def\mr#1{{\mathrm{#1}}}

\def\cH{{\cal H}}

\def\cGv{{\cal G}_\mr{v}}
\def\cGvi{{\cal G}_\mr{v}^{-1}}
\def\cG{{\cal G}}
\def\cGi{{\cal G}^{-1}}

\def\tG{\tilde{\cal G}}
\def\tGi{\tilde{\cal G}^{-1}}
\def\tGiI{\tilde{\cal G}^{I-1}}
\def\tGa{\tilde\Gamma}
\def\tW{\tilde W}
\def\tZ{\tilde Z}
\def\fd#1#2{\frac{\delta#1}{\delta#2}}
\def\fdd#1#2#3{\frac{\delta^2#1}{\delta#2\delta#3}}
\def\fddd#1#2#3#4{\frac{\delta^3#1}{\delta#2\delta#3\delta#4}}

\begin{document}
\title{Renormalization group in the internal space}
\author{J. Polonyi$^{1,2}$ and K. Sailer$^3$}
\affiliation{$^1$ Institute for Theoretical Physics, Louis Pasteur University,
Strasbourg, France}
\affiliation{$^2$ Department of Atomic Physics, Lorand E\"otv\"os University,
Budapest, Hungary}
\affiliation{$^3$ Department for Theoretical Physics, University of Debrecen,
Debrecen, Hungary}
\date{\today}
\begin{abstract}
Renormalization group in the internal space consists of the gradual
change of the coupling constants. Functional evolution equations
corresponding to the change of the mass or the coupling constant
are presented in the framework of a scalar model. The evolution
in the mass which yields the functional generalization of the
Callan-Symanzik equation for the one-particle irreducible
effective action is given in its renormalized, cutoff-independent
form. The evolution of the coupling constant generates an
evolution equation for the two-particle irreducible effective action.
\end{abstract}
\maketitle

\section{Introduction}
This paper presents a modified version of the functional
renormalization group (RG) method aiming at the two-particle irreducible
(2PI) Green functions.
The renormalization group strategy consists of the monitoring
Green functions for the elementary field variables as the
fluctuation modes are turned on gradually \cite{trrg}.
There are two further ingredients which extend this method
into a general purpose, non-perturbative algorithm
to solve quantum field theories. The first is the functional generalization
of the blocking where an infinitesimal fraction of the modes are turned
on or off \cite{wh,polch,morrp,wett}. This reflects the following important change
of strategy. In the original version of the renormalization group
method a finite fraction of the modes are turned on or off
and one relies on the perturbation expansion in obtaining
the change in the dynamics. In order to obtain reliable
results, one should work in high enough order of the
perturbation expansion using the few bare, relevant
vertices. The method of using infinitesimal blocking steps is based
on another systematical approximation where infinitely many
vertices are used in a one-loop renormalization group equation.
The manipulation of a large number of constants can be carried out
by means of their generating functional. This is the reason of looking
for the functional generalization of the renormalization group scheme.
The evolution equation which looks like a one-loop equation
is actually exact as long as it is solved in an infinite dimensional
functional space. Since the general theory of  functional differential
equations is not well developed we have to rely on a truncation of
the evolution equation, the only approximation of the method. It
is based on the projection of the functional differential equation into
a functional subspace. The tailoring of this subspace seems to be an
interesting and a physically better motivated approximation scheme
than the truncation of the perturbative series.

In order to motivate another modification of the original
strategy we recall that the order in which the modes are eliminated
in the renormalization group method,
\be
e^{-\frac{1}{\hbar}S_{n-1}(\phi_1,\cdots,\phi_{n-1})}
=\int d\phi_ne^{-\frac{1}{\hbar}S_n(\phi_1,\cdots,\phi_n)},
\ee
is usually given by means of a scale parameter in the external space of
field theory, the space-time, such as the momentum or energy and the
dynamics is studied as the function of the UV or IR cutoff. In this manner
the short distance modes whose dynamics is rendered more perturbative
by the kinetic energy are eliminated first and the more problematical
nonperturbative modes are treated later when the blocked action is sufficiently
renormalized, $p_1\le p_2\le\cdots\le p_{n-1}\le p_n$.
An alternative to this procedure is the ordering of the modes according to
their scale in the internal space, their amplitude. The sharp cutoff version
of this procedure consists of the study of the $\lambda$-dependence of the
path integral
\be
Z_\lambda=\int_{-\lambda}^\lambda d\phi_1\cdots
\int_{-\lambda}^\lambda d\phi_n e^{-\frac{1}{\hbar}S_n(\phi_1,\cdots,\phi_n)}.
\ee
A more natural, smooth cutoff realization of this strategy is the functional
generalization of the Callan-Symanzik equation \cite{ca,sy} where the mass is
evolved in a relativistic theory \cite{int,sim},
\be
Z=\int D[\phi]e^{-\frac{1}{\hbar}S[\phi]-\frac{\lambda}{2\hbar}\int_x\phi^2_x}.
\ee
Since the mass term dominates the dynamics of free fluctuations
whose wavelength is longer than the inverse mass the two
procedures produce similar scale dependences.

There are in the same time essential differences between the external
and the internal space realizations of the renormalization group.
The most important one is that the
physics is kept fixed along the renormalized trajectory of the external
space method but changes together with the control parameter which
suppresses the fluctuations. This difference arises from the different ways
the UV modes are handled during the evolution. In principle both methods are
similar, their evolution equations look formally equivalent, only the actual
form of the suppression of the modes is different. We may introduce a
straightforward division line between these methods by stating that a
 suppression
mechanism corresponds to the external or internal space scheme if
the corresponding evolution equation is UV finite or divergent.
The internal space methods require the introduction of an UV regulator
which is kept fixed during the evolution and the final products
of the evolution are the bare Green functions, as opposed to the
external space methods where the gliding cutoff renormalizes the
Green functions automatically.

In the first part of this paper we present a modification of the
internal space evolution equation which eliminates this shortcoming
and yields UV finite, renormalized Green functions.
This is achieved by the functional implementation of a
renormalization which is reminiscent of the BPHZ method \cite{bp,hepp,zimm}.
One finds as byproduct a possibility of a Hartree-type optimization
of the unavoidable truncation of the evolution equation.
These issues are studied in the framework of the
effective action for the field variable, the generating functional of the
one-particle irreducible (1PI) Green functions and serve as a
preparation for the second part of this work, the construction of
an internal space renormalization scheme for the computation
of the 2PI Green functions.

This step is made by generalizing the Callan-Symanzik scheme into a
far more flexible procedure by allowing the control parameter,
the mass square in the preceding case, to be replaced by {\em any} parameter
of the dynamics which controls the physics in a differentiable manner. Such
kind of an extension is needed when symmetry considerations require the
presence of soft modes and the suppression of the fluctuations by a mass term
is inconsistent. Actually, the mass controls the {\em amplitude}
of the fluctuations but disregards their {\em nature}. The Gaussian
fluctuations in the absence of interactions are rather peculiar,
their higher moments factories according to Wick's theorem.
The interactions introduce non-factorisable correlations.

It is easy to identify a control parameter which turns the
non-factorisable fluctuations on. In fact, it is just the
coupling constant or a common multiplicative factor in front of
all coupling constants if there are several of them. The evolution equation,
generated by the coupling constant is highly involved \cite{int}.
In fact, the evolution equation contains one-loop integrals as long
as the suppression term which generates the evolution is quadratic
in the field. The complications arising from a quartic suppression
term, $\phi^4$, the two- and three-loop integrals in the evolution equation
reflect the complicated structure of excitations created by the
operator $\phi^4$ in terms of single-particle excitations. One can
simplify the situation, at least formally, by using composite operators
in which the suppression term is again quadratic. The evolution equation
for the effective action for such composite fields will be as simple
as that of the Callan-Symanzik scheme. In the case of the $\phi^4$
scalar theory the composite operator one turns at is obviously $\phi^2$
and one expects that the gradual turning on the interactions should be
possible to realize in a simple manner by following the evolution
of the effective action for the composite field $\phi^2$. But the physics
encapsulated by such an effective action is far more restricted than those
 found
in the effective action for the non-local composite operator
$\phi_x\phi_y$ \cite{cjt,hay}. In fact, the
polarization cloud around a bare particle requires the possibility of
separating the two local, elementary field variables in $\phi^2$ in order to
trace the space-time structure of dressed, physical excitations.
Therefore our proposal, spelled out in the second part is to describe
the structure of vacuum up to two-particle correlations by solving
the evolution equation for the effective action for the propagator.

This scheme is useful if the initial conditions, the effective
action at the initial value of the coupling constant is known.
One should therefore give the initial conditions for vanishing
or infinitesimally weak coupling strength.
The evolution equation will then be integrated until the physical
value of the coupling constant is reached. We shall
investigate in this work the solution arising from the non-interacting
initial condition only. Such a scheme has already been found useful in
performing gauge-free computations in QED \cite{qed},
in constructing systematically the density functional \cite{dft},
and for identifying the localization-localization transition
for disordered systems \cite{loc}.

There are formal similarities between the Schwinger-Dyson
hierarchy of equations and the evolution equation. But the
main advantage of the evolution equation approach is
that it provides a coupled set of equations for the Green functions of the
composite field which are uniquely solvable for a given initial condition
in the perturbative regime.

The organization of the paper is the following. In Sect. \ref{fucasa} a short
description of the functional Callan-Symanzik approach is given, its solution
by iteration and the improvement of the truncated iteration by optimization
is pointed out. Sect. \ref{seveq} presents the derivation of the internal
space RG evolution equation for the 2PI effective action for a family of
composite fields, the bilinear expressions of the elementary field.
The RG equations for the
effective theory are presented in Sect. \ref{expa} after expanding the
2PI effective action around the vacuum. Instead of their numerical solution
an optimized iterative approach for solving the evolution equation of
the 2PI effective action is discussed in Sect. \ref{itersol}, and the
analytic solution found truncating the iteration just after the first step.
The effective theory for the composite field is related to the theory for
the elementary field in Sect. \ref{efthe}.
Finally, the results are summarized in Sect. \ref{su}.

\section{1PI effective action}\label{fucasa}
The simplest realization of the internal space renormalization group method
is the control of the amplitude of the fluctuations by the mass term
which yields a functional generalization of the Callan-Symanzik equation.
As the simplest example we consider the Euclidean scalar model defined by
the bare action
\be\label{hatas}
S[\varphi]=\int_x\left[-\hf\varphi_x\Box_x\varphi_x+\hf \mu^2\varphi^2_x
+\frac{g}{4!}\varphi_x^4\right],
\ee
see Appendix \ref{not} for the notations and conventions.
The generating functional
\be\label{wdef}
e^{\frac1{\hbar} W[j]}=\int\cd{\varphi}e^{-\frac1{\hbar} S[\varphi]
+\frac1{\hbar}\varphi\cdot j}
\ee
is given by a regulated path integral where the UV regulator is not shown
explicitly. It gives rise the effective action
\be
\Gamma[\phi]=-W[j]+j\cdot\phi,~~~\fd{W[j]}{j}=\phi
\ee
satisfying the relations
\be
j= \fd{\Gamma[\phi]}{\phi},~~~~\fdd{W[j]}{j}{j}\cdot\fdd{\Gamma[\phi]}{\phi}{\phi}=1.
\ee

\subsection{Evolution equation}
The dependence of $W[j]$ and $\Gamma[\phi]$ in the mass $\mu^2$, considered
further on as our control parameter, is related,
\be
\partial_{\mu^2} \Gamma[\phi]=-\partial_{\mu^2} W[j],
\ee
and the evolution equation can be obtained by simply bringing
the derivative $\partial_{\mu^2}$ into the functional integration in
Eq. \eq{wdef},
\bea\label{phi4eveq}
\partial_{\mu^2}\Gamma[\phi]&=&\frac{1}{2}\int_x\left[\hbar\fdd{W[j]}{j_x}{j_x}
+\left(\fd{W[j]}{j_x}\right)^2\right]
=\frac{1}{2}\int_x\left[\hbar\left(
\fdd{\Gamma[\phi]}{\phi}{\phi}\right)^{-1}_{x,x}+\phi^2_x\right].
\eea
It is worthwhile mentioning that the
harmless looking step of bringing the partial derivation with respect to
the control parameter of the evolution within the path integral
produces in general highly non-trivial results.
First, it can be considered as the generalization of the
Hellmann-Feynman theorem \cite{hel,fey} for time-dependent processes.
Second, different choices for the control parameter $\lambda$ produce
important, well-known equations: (i) When $\lambda=\varphi_x$, the functional
derivative with respect to one of the integral variables is vanishing.
The stability of the path integral against the shift of the integral
variables produces the Schwinger-Dyson hierarchical equations (c.f. App.
\ref{efv}). (ii) If $\lambda$ is a parameter of a transformation of $\varphi_x$
which preserves at least a part of the action invariant
then the results are the Ward identities.

Here we use the simplest realization of the RG idea
to control the amplitude of the field fluctuations, namely the
control is achieved via the mass term $\lambda=\mu^2$ in a momentum-independent
manner \cite{int} in order to stay as close as possible to the discussion
of the evolution equation for the 2PI effective action
presented below. There are other choices which allow one to control the
amplitude. The most widely used suppression methods
imply strong momentum dependence \cite{wh,polch,morrp,wett}.

The graphical interpretation of  Eq. \eq{phi4eveq}
is obtained by following the Callan-Symanzik strategy. In fact,
the $\mu^2$-dependence of a Feynman graph $F$ with legs amputed
arises from its internal lines, the propagators $G$,
\be\label{casyeeq}
\partial_{\mu^2} F=\sum_p\fd{F}{G_p}\partial_{\mu^2} G_p
=-\sum_p\fd{F}{G_p}S^{(2)-1}_{p}
\partial_{\mu^2} S^{(2)}_{\mu^2,p}S^{(2)-1}_{p}.
\ee
The evolution equation is the application of this chain rule
for the 1PI graphs in the framework of the skeleton expansion.

Contrary to the strategy of the Callan-Symanzik equation where
one expands around the massless, critical system our initial condition
is supposed to be in the perturbative regime. Therefore we choose
$\mu_0^2\gg\Lambda^2$ where $\Lambda$ denotes the UV cutoff
as initial condition because the quantum fluctuations are suppressed
and the tree-level approximation to the effective action,
\be\label{casyic}
\Gamma_\mr{in}[\phi]=\int_x\left[-\hf\phi_x\Box_x\phi_x
+\frac{\mu_0^2}{2}\phi^2_x+\frac{g}{4!}\phi_x^4\right]
\ee
is applicable.

\subsection{Renormalization}
There are two related reasons to perform further adjustments on the
evolution equation \eq{phi4eveq}. One is that the control parameter
used in the internal space evolution does not regulate the UV divergences.
As a result Eq. \eq{phi4eveq} has to be first regulated and the resulting
effective action must be renormalized in order
to extract Green functions which correspond to cutoff independent
scales. Another, more technical looking complication is that
the integration of the evolution equation \eq{phi4eveq},
starting with the initial condition \eq{casyic}, generates more and more
complicated $\phi$-dependence in the effective action as
$\mu^2$ is decreased. The two issues are related because once
the effective action is expressed in terms of the renormalized
quantities it is supposed to contain slower, i.e. easier to truncate
dependence in the field.

The unavoidable truncation of the evolution equation
should be as harmless as possible from the point of view of the terms
appearing in the inverse of the right hand side of the evolution equation
\eq{phi4eveq}.
There are two different directions to follow in the search of optimization.
One is the appropriate choice of the dependence on the control parameter
of the evolution in the Lagrangian. Because all we want to achieve is an
interpolation between a simple, calculable initial
condition and the true system only, different possibilities
might be available to introduce the control parameter.
This avenue has already been explored in the
context of optimizing the blocking transformation
in the traditional realization of the renormalization group
method in external space \cite{optsb,optdl}. We cannot follow this direction here,
having already chosen the mass as control parameter because of other reason.
The other dimension of the optimization is the tailoring of the
functional form of the effective action.
One should recall in this respect that the magnitude of the
variable $\phi$ has nothing to do with the strength of the
quantum fluctuations since $\phi$ is a book-keeping device only
in defining the generating functional for the 1PI Green functions.
Nevertheless $\phi$ is supposed to be small in an indirect manner
for an optimized scheme because the more complicated terms
of the effective action tend to be of higher order in $\phi$.
More precisely, the omission of the terms $\ord{\phi^{2n}}$
leads to ignoring the $n$-particle correlations.

We shall try to slow down the $\phi$-dependence of the effective action
by the appropriate truncation of the inverse on the right
hand side of the evolution equation. The inverse will be written as a series
where terms appear in the order of increasing complexity
in such a manner that the more complicated terms are as small
as possible. The tailoring of the "free" inverse propagator around
which the inverse is expanded is reminiscent of the Hartree
approximation when one adds a mass term to the free Lagrangian,
subtracts it from the interaction part, and optimizes the
perturbation expansion with respect to this rearrangement of the
Lagrangian. But one should notice that the same strategy is followed in
establishing the renormalized perturbative series, too. Therefore we
separate off local counterterms, $\Gamma_{CT}[\phi]$,
from the effective action,
\be\label{sepphi4}
\Gamma[\phi]=\Gamma_R[\phi]+\Gamma_{CT}[\phi]
\ee
in such a manner that the renormalized effective action,
$\Gamma_R[\phi]$, converges as the cutoff is removed.
While this separation has no effect on the exact solution, it will
influence the solution of the truncated evolution equation.
The condition $\Lambda\ll\mu^2_0$ clearly prevents us to send the cutoff to
infinity. But an approximate cutoff independence of $\Gamma_R[\phi]$
can still be tested for $\mu^2\ll\Lambda^2\ll\mu^2_0$ for a
sufficiently large but finite $\mu^2_0$.

The functional Taylor expansion involving the renormalized 1PI Green
functions $\Gamma_{n\;x_1,\ldots,x_n}$ and their counterterms,
$\Gamma_{CT~n\;x_1,\ldots,x_n}$, will be used,
\bea\label{ftexp}
\Gamma_R[\phi]=\sum_{n=0}^\infty\frac{1}{(2n)!}
\Gamma_{2n\;x_1,\ldots,x_{2n}}\phi_{x_1}\cdots\phi_{x_{2n}},&~~&
\Gamma_{CT}[\phi]=\sum_{n=0}^\infty\frac{1}{(2n)!}
\Gamma_{CT~2n\;x_1,\ldots,x_{2n}}\phi_{x_1}\cdots\phi_{x_{2n}},
\eea
where double indices are integrated over, c.f. Appendix \ref{not}.
The evolution equation
\be\label{evegyfeleri}
\partial_{\mu^2}\Gamma_R[\phi]+\partial_{\mu^2}\Gamma_{CT}[\phi]
=\frac{\hbar}{2}\tr\left(\fdd{\Gamma_R[\phi]}{\phi}{\phi}
+\fdd{\Gamma_{CT}[\phi]}{\phi}{\phi}\right)^{-1}+\hf\phi\cdot\phi
\ee
is then written by means of expanding the inverse on the right
hand side around the renormalized propagator, $G_R=\Gamma_2^{-1}$, and
by splitting the resulting series into the sum of two pieces in $d$ dimensions,
\bea\label{evegyeri}
\partial_{\mu^2}\Gamma_R[\phi]&=&\frac{\hbar}{2}\sum_{m=[\frac{d}{2}]+1}^\infty
(-1)^m\tr\left[G_R\cdot\left(\sum_{n=2}^\infty\frac{1}{(2n-2)!}
\Gamma_{2n\;x_3,\ldots,x_{2n}}\phi_{x_3}\cdots\phi_{x_{2n}}\cdot
G_R\right)^m\right]+\hf\phi\cdot\phi,
\eea
and
\bea\label{evketi}
\partial_{\mu^2}\Gamma_{CT}[\phi]&=&
\frac{\hbar}{2}\tr\left(\fdd{\Gamma_R[\phi]}{\phi}{\phi}
+\fdd{\Gamma_{CT}[\phi]}{\phi}{\phi}\right)^{-1}\nn
&&-\frac{\hbar}{2}\sum_{m=[\frac{d}{2}]+1}^\infty(-1)^m\tr\left[G_R\cdot
\left(\sum_{n=2}^\infty\frac{1}{(2n-2)!}
\Gamma_{2n\;x_3,\ldots,x_{2n}}\phi_{x_3}\cdots\phi_{x_{2n}}\cdot
G_R\right)^m\right],
\eea
where the first two space-time indices of $\Gamma_n$ were suppressed.
This split is made in such a manner that the renormalized part remains
"finite", i.e. approximately cutoff independent at energy scales
$p$ as long as $p^2\ll\Lambda^2\ll\mu_0^2$. The expansion of the inverse
on the right hand side of Eq. \eq{evketi} around $G_R$,
\bea
\partial_{\mu^2}\Gamma_{CT}[\phi]
&=&\frac{\hbar}{2}\sum_{m=0}^\infty (-1)^m\tr\left\{G_R\left[\left(
\Gamma_{CT~2}+\sum_{n=2}^\infty\frac{1}{(2n-2)!}(\Gamma_{2n~x_3,\ldots,x_{2n}}
+\Gamma_{CT~2n~x_3,\ldots,x_{2n}})\phi_{x_3}\cdots\phi_{x_{2n}}
\right)G_R\right]^m\right\}\nn
&&-\frac{\hbar}{2}\sum_{m=[\frac{d}{2}]+1}^\infty(-1)^m\tr\left\{G_R\left[\sum_{n=2}^\infty
\frac{1}{(2n-2)!}\Gamma_{2n~x_3,\ldots,x_{2n}}\phi_{x_3}\cdots\phi_{x_{2n}}
G_R\right]^m\right\},
\eea
shows that each contribution to $\Gamma_{CT}[\phi]$ is UV divergent.

The identification of the coefficients of the first few powers of
$\phi$ on both sides gives
\bea\label{etaeveq}
\partial_{\mu^2}\Gamma_0&=&0,\nn
\partial_{\mu^2}\Gamma_{2,x_1,x_2}&=&\delta_{x_1,x_2},\nn
\partial_{\mu^2}\Gamma_{4,x_1,x_2,x_3,x_4}&=&0,\nn
\partial_{\mu^2}\Gamma_{6,x_1,x_2,x_3,x_4,x_5,x_6}&=&-\frac{\hbar}{2\cdot2^3}
\sum_{\pi\in S_6}\tr[G_R\Gamma_{4~x_{\pi(1)},x_{\pi(2)}}G_R
\Gamma_{4~x_{\pi(3)},x_{\pi(4)}}G_R\Gamma_{4~x_{\pi(5)},x_{\pi(6)}}G_R],
\eea
and
\bea
\partial_{\mu^2}\Gamma_{CT~0}&=&\frac{\hbar}{2}\sum_{m=0}^\infty(-1)^m
\tr[G_R(\Gamma_{CT~2}G_R)^m]
=\frac{\hbar}{2}\tr(\Gamma_2+\Gamma_{CT~2})^{-1},\nn
\partial_{\mu^2}\Gamma_{CT~2,x_1,x_2}&=&\frac{\hbar}{4}\sum_{m=1}^\infty(-1)^m
\sum_{\pi\in S_2}\sum_{j=1}^m\tr\left[G_R(\Gamma_{CT~2}G_R)^{j-1}
G_R(\Gamma_{CT~2}G_R)^{m-j}(\Gamma_{4}+\Gamma_{CT~4})_{x_{\pi(1)},x_{\pi(2)}}
G_R\right],\nn
\partial_{\mu^2}\Gamma_{CT~4,x_1,x_2,x_3,x_4}&=&\frac{\hbar}{8}\sum_{m=2}^\infty(-1)^m
\sum_{\pi\in S_4}\sum_{j=1}^{m-2}\sum_{k=0}^{m-j-2}
\tr\bigl[G_R(\Gamma_{CT~2}G_R)^{m-j-k-2}
(\Gamma_{4}+\Gamma_{CT~4})_{x_{\pi(1)},x_{\pi(2)}}G_R\nn
&&\times(\Gamma_{CT~2}G_R)^{j}(\Gamma_{4}+\Gamma_{CT~4})_{x_{\pi(3)},x_{\pi(4)}}
G_R(\Gamma_{CT~2}G_R)^{k}\bigr]\nn
&&+\frac{\hbar}{48}\sum_{m=1}^\infty(-1)^m\sum_{\pi\in S_4}\sum_{j=1}^m\nn
&&\times\tr\left[G_R(\Gamma_{CT~2}G_R)^{j-1}G_R(\Gamma_{CT~2}G_R)^{m-j}
(\Gamma_{6}+\Gamma_{CT~6})_{x_{\pi(1)},x_{\pi(2)},x_{\pi(3)},x_{\pi(4)}}G_R\right],
\eea
for the first few coefficient functions.

\subsection{Iterative solution}
It is instructive to imagine the solution of the evolution equation
as the fixed point of the iteration $\Gamma^{[N-1]}[\phi]\to\Gamma^{[N]}[\phi]$,
\be\label{evint}
\Gamma^{[N+1]}_{\mu^2}[\phi]=\Gamma^{[N]}_{\mu_0^2}[\phi]
+\frac{\hbar}{2}\int_{\mu_0^2}^{\mu^2}d\tilde\mu^2
\tr\left(\fdd{\Gamma^{[N]}_{\tilde\mu^2}[\phi]}{\phi}{\phi}\right)^{-1}
+\frac{\mu^2-\mu_0^2}{2}\phi\cdot\phi.
\ee
The speed of convergence of the iterations depends on the choice of
the starting point. Let us assume in the spirit of the perturbation expansion
that the iteration reaches at least a fixed point when the
tree-level starting point, given by Eq. \eq{casyic}, is used.
The iteration in Eq. \eq{evint} becomes well-defined at each step
only after imposing
the initial condition \eq{casyic} together with $\Gamma_{CT~\mu_0^2}[\phi]=0$
 for the
$\mu^2$-integration, i.e. the dressing is
assumed to be completely suppressed at $\mu_0^2$.

For the sake of simplicity, we apply the local potential approximation, i.e.
we restrict the field variable to be homogeneous, $\phi_x\to\Phi$, in
the effective action which will be written as
\be\label{udef}
\Gamma_{\mu^2}[\Phi]=V[U_{R~\mu^2}(\Phi)+U_{CT~\mu^2}(\Phi)].
\ee
The starting point is a $\mu^2$-independent potential,
\bea
U^{[0]}_{R~\mu^2}(\Phi)&=&U_{\mu_0^2}(\Phi)
=\frac{\mu_0^2}{2}\Phi^2+\frac{g}{4!}\Phi^4,\nn
U^{[0]}_{CT~\mu^2}(\Phi)&=&0,
\eea
and the iteration \eq{evint} becomes an integral equation,
\bea\label{integy}
U^{[N+1]}_{R~\mu^2}(\Phi)&=&\frac{\mu^2}{2}\Phi^2+\frac{g}{4!}\Phi^4
+B_{R~\mu_0^2\to\mu^2}[U_{R~\mu^2}^{(N)}],\nn
U^{[N+1]}_{CT~\mu^2}(\Phi)&=&U_{CT~\mu_0^2}^{[N]}
+B_{CT~\mu_0^2\to\mu^2}[U_{R~\mu^2}^{[N]},U_{CT~\mu^2}^{[N]}]
\eea
for the local potentials which is written in $d=4$ as
\bea\label{imap}
B_{R~\mu_0^2\to\mu^2}[U_{R~\mu^2}]&=&-\frac{\hbar}{2}\int_{\mu^2}^{\mu_0^2}
d\tilde\mu^2\int_p\left[\frac{1}{p^2+U''_{R~\tilde\mu^2}(\Phi)}
-\left(1-V''_{R~\tilde\mu^2}(\Phi)G_{R~\tilde\mu^2}
+[V''_{R~\tilde\mu^2}(\Phi)G_{R~\tilde\mu^2}]^2\right)G_{R~\tilde\mu^2}\right]\nn
&=&-\frac{\hbar}{2}\int_{\mu^2}^{\mu_0^2}d\tilde\mu^2\sum_{m=3}^\infty\int_p
[(V''_{R~\tilde\mu^2}(\Phi)G_{R~\tilde\mu^2}]^mG_{R~\tilde\mu^2},\nn
B_{CT~\mu_0^2\to\mu^2}[U_{R~\mu^2},U_{CT~\mu^2}]&=&-\frac{\hbar}{2}
\int_{\mu^2}^{\mu_0^2}
d\tilde\mu^2\int_p\biggl[\frac{1}{p^2+U''_{R~\tilde\mu^2}(\Phi)
+U''_{CT~\tilde\mu^2}(\Phi)}
-\frac{1}{p^2+U''_{R~\tilde\mu^2}(\Phi)}\nn
&&+\left(1-V''_{R~\tilde\mu^2}(\Phi)G_{R~\tilde\mu^2}
+[V''_{R~\tilde\mu^2}(\Phi)G_{R~\tilde\mu^2}]^2\right)G_{R~\tilde\mu^2}\biggr]\nn
&=&-\frac{\hbar}{2}\int_{\mu^2}^{\mu_0^2}d\tilde\mu^2\int_p
\left[1-V''_{R~\tilde\mu^2}(\Phi)G_{R~\tilde\mu^2}
+[V''_{R~\tilde\mu^2}(\Phi)G_{R~\tilde\mu^2}]^2\right]G_{R~\tilde\mu^2}\nn
&&-\frac{\hbar}{2}\int_{\mu^2}^{\mu_0^2}d\tilde\mu^2\int_p
\sum_{m=1}^\infty(-1)^m\frac{1}{p^2+U''_{R~\tilde\mu^2}(\Phi)}
\left(\frac{U''_{CT~\tilde\mu^2}(\Phi)}
{p^2+U''_{R~\tilde\mu^2}(\Phi)}\right)^m,\nn
\eea
with the help of the renormalized propagator
\be
G_{R~\tilde\mu^2}=\frac{1}{p^2+U''_{R~\tilde\mu^2}(0)}
\ee
and the subtracted potential
\be
V_{R~\mu^2}(\Phi)=U_{R~\mu^2}(\Phi)-\hf U_{R~\mu^2}''(0)\Phi^2.
\ee

The multiplicative factors $\hbar$ in front of the one-loop integral on the
right hand sides indicate that each iteration resums
a successive order of the loop-expansion within the local potential
approximation. The result of the first iteration is
called independent mode approximation because the quantum fluctuations are
treated independently. The main simplification in computing in this order
comes from the fact that the initial action is independent of $\mu^2$,
\bea\label{uctima}
U^{[1]}_{R~\mu^2}(\Phi)&=&\frac{\mu^2}{2}\Phi^2+\frac{g}{4!}\Phi^4
+U^{1l}(\Phi;\mu^2,g)- U^{1l}(\Phi;\mu_0^2,g)
+\frac{\hbar}{2}\int^{\mu_0^2}_{\mu^2}d\tilde\mu^2\sum_{n=0}^{2}
(-1)^n\int_p\left(\frac{\frac{g}{2}\Phi^2}{p^2+\tilde\mu^2}\right)^n
\frac{1}{p^2+\tilde\mu^2}\nn
&=&\frac{\mu^2}{2}\Phi^2+\frac{g}{4!}\Phi^4
+U^{1l}_R(\Phi;\mu^2,g)-U^{1l}_R(\Phi;\mu^2_0,g)\nn
U^{[1]}_{CT~\mu^2}(\Phi)&=&-\frac{\hbar}{2}\int_{\mu^2}^{\mu_0^2}
d\tilde\mu^2\sum_{n=0}^{2}(-1)^n\int_p
\left(\frac{\frac{g}{2}\Phi^2}{p^2+\tilde\mu^2}\right)^n
\frac{1}{p^2+\tilde\mu^2}\nn
&=&\frac{\hbar}{2}\int_p\left[\ln\frac{p^2+\mu^2}{p^2+\mu_0^2}
+\frac{g}{2}\Phi^2\left(\frac{1}{p^2+\mu^2}-\frac{1}{p^2+\mu_0^2}\right)
-\hf\biggl(\frac{g}{2}\Phi^2\biggr)^2\left(\frac{1}{(p^2+\mu^2)^2}
-\frac{1}{(p^2+\mu_0^2)^2}\right)\right],
\eea
where
\be
U^{1l}(\Phi;\mu^2,g)=\frac{\hbar}{2}\int_p\ln\left(p^2+\mu^2+\frac{g}{2}\Phi^2\right)
\ee
stands for the one-loop effective potential of the model of mass $\mu$
and coupling constant $g$ and
\be
U^{1l}_R(\Phi;\mu^2,g)=U^{1l}(\Phi;\mu^2,g)-\frac{\hbar}{2}
\int_p\left[\ln(p^2+\mu^2)+\frac{g}{2}\Phi^2\frac{1}{p^2+\mu^2}
-\hf\left(\frac{\frac{g}{2}\Phi^2}{p^2+\mu^2}\right)^2\right]
\ee
represents its UV finite, renormalized part. When the limit $\mu_0^2\to\infty$
is made then the $\ord{\Phi^2/\mu_0^2}$ terms become negligible and
we find the standard result,
\bea
U^{[1]}_{R~\mu^2}(\Phi)&=&\frac{\mu^2}{2}\Phi^2+\frac{g}{4!}\Phi^4
+U^{1l}_R(\Phi;\mu^2,g),\nn
U^{[1]}_{CT~\mu^2}(\Phi)&=&\frac{\hbar}{2}\int_p\left[\ln(p^2+\mu^2)
+\frac{g}{2}\Phi^2\frac{1}{p^2+\mu^2}-\hf\left(\frac{g}{2}\Phi^2\right)^2
\frac{1}{(p^2+\mu^2)^2}\right],
\eea
up to a constant in the renormalized potential. It is easy to see that
the subsequent iterations preserve the UV finiteness of $U_R(\Phi)$.

It is worthwhile mentioning that the split of the iteration of Eq. \eq{evint}
into a renormalized and a counterterm part,
\bea
\partial_{\mu^2}\Gamma_R^{[N+1]}[\phi]
&=&\frac{\hbar}{2}\sum_{m=[\frac{d}{2}]+1}^\infty
(-1)^m\tr\left[G_R\cdot\left(\sum_{n=2}^\infty\frac{1}{(2n-2)!}
\Gamma_{2n\;x_3,\ldots,x_{2n}}^{[N]}\phi_{x_3}\cdots\phi_{x_{2n}}\cdot
G_R\right)^m\right]+\hf\phi\cdot\phi,
\eea
and
\bea\label{ellent}
\partial_{\mu^2}\Gamma_{CT}^{[N+1]}[\phi]
&=&\frac{\hbar}{2}\sum_{m=0}^\infty (-1)^m\tr G_R\left[\left(
\Gamma_{CT~2}^{[N]}+\sum_{n=2}^\infty\frac{1}{(2n-2)!}
(\Gamma_{2n~x_3,\ldots,x_{2n}}^{[N]}
+\Gamma_{CT~2n~x_3,\ldots,x_{2n}}^{[N]})\phi_{x_3}\cdots\phi_{x_{2n}}
\right)G_R\right]^m\nn
&&-\frac{\hbar}{2}\sum_{m=[\frac{d}{2}]+1}^\infty(-1)^m\tr
G_R\left[\sum_{n=2}^\infty\frac{1}{(2n-2)!}\Gamma_{2n~x_3,\ldots,x_{2n}}^{[N]}
\phi_{x_3}\cdots\phi_{x_{2n}}G_R\right]^m,
\eea
realizes a particular BPHZ renormalization scheme. In fact, $\Gamma_R^{[N]}[\phi]$
is a finite functional for each $N$ and the counterterms in
$\Gamma_{CT}^{[N]}[\phi]$ correspond to the full subtraction of the
overall divergences at each order $N$ which are defined by the
expansion of the integrand of the loop integral, the trace on the
right hand side of Eq. \eq{ellent}.

\section{2PI effective action}\label{seveq}
Our goal in this section is to derive the evolution equation for the 2PI
effective action for the model defined by the action \eq{hatas}. This is 
achieved by
generalizing the standard procedure, presented briefly in section \ref{fucasa},
for the bilocal field $\varphi_x\varphi_y$. For this end
we introduce the generating functional
\be\label{www}
e^{\frac1{\hbar}\tW[J]}=\int\cd{\varphi}e^{-\frac1{\hbar}S[\varphi]
+\frac1{2\hbar}\varphi\cdot J\cdot\varphi},
\ee
together with its Legendre transform, the effective action
for the two-point function,
\be\label{oda}
\cG_{x,y}=\la T(\varphi_x\varphi_y)\ra=\fd{\tW[J]}{\hf J_{x,y}},
\ee
which is given by
\be
\tilde\Gamma[\cG]=-\tW[J]+\hf\tr\cG^t\cdot J.
\ee
The inverse Legendre transformation is based on the relation
\be\label{vissza}
\fd{\tilde\Gamma[\cG]}{\cG_{x,y}}=\hf J_{x,y}.
\ee
Since the source $J_{x,y}$ is symmetrical, so is $\cG_{x,y}$.
Another important symmetry, translational invariance is supposed
to be recovered in the limit $J\to0$, therefore the propagator in
the vacuum is taken to be translation invariant,
$\la T(\varphi_x\varphi_y)\ra_{J=0}=\cG_{\mr{v}~x,y}=\cG_{\mr{v}~x-y}$.
For the sake of simplicity we shall assume
that the bare action is always chosen in such a manner that there
is no condensation, $\la\varphi_x\ra_{J=0}=0$, as in Section \ref{fucasa}.

The relation
\be
I_{(x_1,x_2),(y_1,y_2)}=\hf(\delta_{x_1,y_1}\delta_{x_2,y_2}
+\delta_{x_1,y_2}\delta_{x_2,y_1})
=\fd{\cG_{x_1,x_2}}{\cG_{y_1,y_2}},
\ee
cf. Eq. \eq{gfder}, can be used to establish the identity
\be\label{inverzwg}
\fdd{\tW[J]}{\hf J}{\hf J}
=\left[\fdd{\tilde\Gamma[\cG]}{\cG}{\cG}\right]^{-1}: I
=\left[\fdd{\tilde\Gamma[\cG]}{\cG}{\cG}\right]^{-1I},
\ee
where the inverse $A^{-1I}\equiv A^{-1}: I$ with respect to
the operator $I$ was introduced.
The matrix $I$ acts on the two-particle subspace by projecting
into the symmetrical subspace and plays the
role of the identity operator for indistinguishable particles.

\subsection{Evolution equation}
We shall use  the coupling constant $g$ to control the RG evolution
in internal space and let it go from zero to its physical value.
The corresponding evolution of $\tilde\Gamma[\cG]$ is governed
by the evolution equation
\bea\label{evolk}
\partial_g \tilde\Gamma[\cG]&=&-\partial_g \tW[J]
=e^{-\frac1{\hbar}\tW[J]}\int\cd{\varphi}\partial_g S[\varphi]
e^{-\frac1{\hbar}S[\varphi]+\frac1{2\hbar}\varphi\cdot J\cdot\varphi}
=\frac{1}{4!}\int_x\left[\hbar
\left(\fdd{\tilde\Gamma[\cG]}{\cG}{\cG}\right)^{-1I}_{(x,x),(x,x)}
+(\cG_{x,x})^2\right].
\eea
It is worthwhile mentioning that the onset of the condensation is signaled
by the vanishing of the restoring forces to the equilibrium
position, acting on the quantum fluctuations, i.e. by $\cG^{-1}_p=0$.
Therefore, the exclusion of the condensation induces the
restriction $\cG^{-1}_p\not=0$.

It is easy to understand the evolution equation for the functional $\tW[J]$
graphically. In fact, let us assume the form
\be\label{wftser}
\tW[J]=\sum_{n=0}^\infty\frac{1}{n!}
\tW_{n\;X_1,\ldots,X_n}J_{X_1}\cdots J_{X_n},
\ee
where the repeated double-indices $X_j=(x_j,y_j)$ are integrated over
and the evolution equation reads as
\bea\label{grjueveq}
&&\sum_{n=0}^\infty\frac{1}{n!}\partial_g
\tW_{n\;X_1,\ldots,X_n}J_{X_1}\cdots J_{X_n}
=\sum_{n=0}^\infty\frac{1}{n!}\frac{\hbar}{3!}\int_x
\tW_{n+2\;(x,x),(x,x),X_1,\ldots,X_n}J_{X_1}\cdots J_{X_n}\nn
&&~~~~+\frac{1}{3!}\int_x\sum_{m,n=0}^\infty\frac{1}{m!n!}
\tW_{m+1\;(x,x),X_1,\ldots,X_m}J_{X_1}\cdots J_{X_m}
\tW_{n+1\;(x,x),Y_1,\ldots,Y_n}J_{Y_1}\cdots J_{Y_n}.
\eea
Consider the contribution $\ord{J^k}$ on both sides.
On the left hand side we find $\partial_g$ acting on the
Green function with $k$ pairs of field variables which is
connected as far as the cutting of two particle lines are concerned.
This Green function is the sum of Feynman graphs and the dependence
on $g$ appears through the factors $ g$ multiplying the
vertices. Therefore, the right hand side is supposed to be the sum
of these contributions. We turn now the four internal lines
attached to the vertex on which $\partial_g$ acts into
external lines and consider the left-over graph.
The first and the second terms on the right hand side stand
for graphs whose left-over graph remains two-particle connected
or falls into two disconnected components, respectively.

Motivated by the Hartree-type optimization of the truncated functional
Callan-Symanzik equation presented in section \ref{fucasa} we split off
a quadratic term from the effective action,
\be\label{separ}
\tilde\Gamma[\cG]=\Gamma[\cG]+\frac{1}{2}\alpha_{X_1,X_2}\cG_{X_1}\cG_{X_2},
\ee
where $\alpha$ denotes a $g-$dependent coefficient function which is
vanishing for $g=0$. The corresponding evolution equation is
\be\label{egyszev}
\partial_g\Gamma[\cG]=\frac{\hbar}{4!}\int_x\left[
\fdd{\Gamma[\cG]}{\cG}{\cG}+\alpha\right]^{-1I}_{(x,x),(x,x)}
+\cG:\left(\frac{1}{4!}L-\hf\partial_g\alpha\right):\cG,
\ee
with $L_{(x,y),(u,v)}=\delta_{x,y}\delta_{u,v}\delta_{x,u}$.
It is furthermore useful to separate the effects of interactions in the
effective action by means of the parametrization
\be\label{nintpar}
\Gamma[\cG]=\Gamma^{free}[\cG]+U[\cG],
\ee
where the first  term corresponds to the free theory, cf. Appendix \ref{olefac},
\be\label{initk}
\Gamma^{free}[\cG]=\frac{\hbar}{2}\tr\left(\cG\cdot\fdd{S[0]}{\phi}{\phi}\right)
-\frac{\hbar}{2}\tr\ln\cG,
\ee
and the `potential' $U[\cG]$ stands for the effects of the interactions.
The initial condition, $U[\cG]=0$, for the evolution equation is
imposed at $g=0$.

\subsection{Expansion around the vacuum}\label{expa}
We return now to the evolution equation  \eq{egyszev} which
cannot be solved exactly.
Its usefulness depends on the possibility of finding a good enough
truncation, a projection into a restricted functional space. This latter
should be, on the one hand, simple enough that the truncated equations
can be solved either analytically or numerically and on the other hand,
rich enough to incorporate the important effective vertices of
the theory. The natural procedure is  the functional
Taylor expansion of the type Eq. \eq{wftser} for $U[\cG]$,
truncated at order $N$, i.e. we retain the $N$-particle effective
interactions. In order to render the Taylor series faster converging
and to recover the 2PI vertex functions the Taylor expansion will be
organized around the vacuum by splitting
the propagator as $\cG=\cGv+\cH$ where $\cGv$ is the propagator
in the vacuum and writing the effective action as
\be\label{gans}
\Gamma[\cH]=\Gamma^{free}[\cGv+\cH]+\gamma[\cH],
\ee
with
\be
\gamma(\cH)=U[\cGv+\cH]=\sum_{n=0}^N\frac{1}{n!}
\gamma_{n\;X_1,\ldots,X_n}\cH_{X_1}\cdots\cH_{X_n}.
\ee
The coefficient functions are symmetrical
with respect to the exchange of any pairs of their variables,
$\gamma_{n\ldots,X_j,\ldots,X_k,\ldots}
=\gamma_{n,\ldots,X_k,\ldots,X_j,\ldots}$ and
are translation invariant because
they correspond to the functional Taylor expansion around
the translation invariant vacuum. This justifies the notation
\be
\gamma_{n,(p_1,q_1),\ldots,(p_n,q_n)}=\delta_{p_1+q_1+\cdots+p_n+q_n,0}
\hat\gamma_{n,(p_1,q_1),\ldots,(p_n,q_n)}
\ee
in momentum space. The functional Taylor expansion for the full
effective action thus reads as
\be
\Gamma[\cH]=\sum_{n=0}^\infty\frac{1}{n!}
\Gamma_{n~X_1,\cdots,X_n}\cH_{X_1}\cdots\cH_{X_n},
\ee
where the first few coefficients are
\bea
\Gamma_0&=&\frac{\hbar}{2}\tr\left(\cGv\cdot\fdd{S[0]}{\phi}{\phi}\right)
-\frac{\hbar}{2}\tr\ln\cGv+\hat\gamma_0\nn
\Gamma_{1;(x,y)}&=&\frac{\hbar}{2}\fdd{S[0]}{\phi_y}{\phi_x}
-\frac{\hbar}{2}\cGi_{\mr{v}~y,x}
+\hat\gamma_{1;(x,y)},\nn
\Gamma_{2,X_1,X_2}&=&\tGiI_{\mr{v}~X_1,X_2}
+\hat\gamma_{2,X_1,X_2},\nn
\Gamma_{3;(x_1,y_1),(x_2,y_2),(x_3,y_3)}
&=&-\frac{\hbar}{2\cdot3!}\sum_{\pi\in S_3}
\cGi_{\mr{v}~(y_{\pi(1)},x_{\pi(2)})}
\cGi_{\mr{v}~(y_{\pi(2)},x_{\pi(3)})}
\cGi_{\mr{v}~(y_{\pi(3)},x_{\pi(1)})}
+\hat\gamma_{3;(x_1,y_1),(x_2,y_2),(x_3,y_3)},\nn
\Gamma_{4;(x_1,y_1),(x_2,y_2),(x_3,y_3),(x_4,y_4)}
&=&\frac{\hbar}{2\cdot4!}\sum_{\pi\in S_4}
\cGi_{\mr{v}~(y_{\pi(1)},x_{\pi(2)})}\cGi_{\mr{v}~(y_{\pi(2)},x_{\pi(3)})}
\cGi_{\mr{v}~(y_{\pi(3)},x_{\pi(4)})}\cGi_{\mr{v}~(y_{\pi(4)},x_{\pi(1)})}\nn
&&+\hat\gamma_{4;(x_1,y_1),(x_2,y_2),(x_3,y_3),(x_4,y_4)},
\eea
where
\be
\tGiI_{(x,y),(u,v)}=\tGi_{(x,y),(u,v)}
=\fdd{\Gamma^{free}[\cG]}{\cG_{x,y}}{\cG_{u,v}}
=\frac{\hbar}{4}(\cG_{v,x}^{-1}\cG_{y,u}^{-1}+\cG_{v,y}^{-1}\cG_{x,u}^{-1}),
\ee
see Appendix \ref{fourier}. The propagator in the vacuum satisfies the equation
\be
\cGvi=\fdd{S[0]}{\phi}{\phi}+\Sigma,
\ee
where the self energy is given by
\be\label{sajatener}
\Sigma=2\alpha:\cG_\mr{v}+2\hat\gamma_1.
\ee

The evolution equation can now be written as
\bea
\partial_g\Gamma&=&\frac{\hbar}{4!}\sum_{m=0}^\infty(-1)^m\int_x
\left\{K:\left[\left(\sum_{n=3}^\infty\frac{1}{(n-2)!}\Gamma_{n;Z_3,\cdots,Z_n}
\cH_{Z_3}\cdots\cH_{Z_n} + \alpha\right):K\right]^m\right\}_{(x,x),(x,x)}\nn
&&+\cG:\left(\frac{1}{4!}L-\partial_g\alpha\right):\cG
+\sum_{n=1}^\infty\frac{1}{(n-1)!}\Gamma_{n;Z_1,Z_2,\cdots,Z_n}
\cH_{Z_2}\cdots\cH_{Z_n}\partial_g\cG_{\mr{v}~Z_1}
+ \cG_{\mr{v}}:\alpha:
\partial_g \cG_{\mr{v}},
\eea
by using $\cH$ as an independent variable and suppressing
the first two double-indices of the coefficient functions $\Gamma_n$. Here
\be
K^{-1}_{X,Y}=\tGiI_{\mr{v}~X,Y}+\hat\gamma_{2,X,Y}
\ee
stands for the inverse of the two-particle propagator and the separated
$\ord{\cG^2}$ term in the effective action is treated by the expansion.
The identification of the coefficients of different powers of $\cH$
on the two sides up to $\ord{\cH^2}$ yields
\bea
\partial_g\hat\gamma_0&=&\frac{\hbar}{4!}\int_xK_{(x,x),(x,x)}
+\frac{\hbar}{4!}\int_x\left[K:(\alpha:K)^m\right]_{(x,x),(x,x)}
+\cG_\mr{v}:\left(\frac{1}{4!}L-\frac{1}{2}\partial_g\alpha\right):\cG_\mr{v}
+(\Gamma_1+\cG_\mr{v}:\alpha):\partial_g\cG_\mr{v},\nn
\partial_g\gamma_{1,X}&=&-\frac{\hbar}{4!}\hat K_{Z_1,Z_2}\Gamma_{3;Z_2,Z_1,X}
+\frac{\hbar}{4!}\tr\hat K:(\alpha:K:\Gamma_{3;X}+\Gamma_{3;X}:K:\alpha)\nn
&&+\frac{1}{3!} L_{X,Z}\cG_{\mr{v}~Z}-\partial_g\alpha_{X,Z}\cG_{\mr{v}~Z}
+(\Gamma_{2;X,Z}+\alpha_{X,Z})\partial_g\cG_{\mr{v}~Z},\nn
\partial_g\hat\gamma_{2,X_1,X_2}&=&\frac{\hbar}{4!}
\sum_{\pi\in S_2}\hat K_{Z_1,Z_2}\left(
\Gamma_{3;Z_2,Z_3,X_{\pi(1)}}K_{Z_3,Z_4}\Gamma_{3;Z_4,Z_1,X_{\pi(2)}}
-\Gamma_{4;Z_2,Z_1,X_{\pi(1)},X_{\pi(2)}}\right)\nn
&&-\partial_g\alpha_{X_1,X_2}+\Gamma_{3,X_1,X_2,Z}\partial_g\cG_{\mr{v}~Z},
\eea
with
\be
\hat K_{X,Y}=\int_uK_{X,(u,u)}K_{(u,u),Y} .
\ee

\subsection{Iterative solution}\label{itersol}
We follow the general strategy of Section \ref{fucasa} and rewrite
the evolution equation \eq{egyszev} as integral equation, used
for iteration,
\be\label{ketiter}
\Gamma_g^{[N+1]}[\cG]=\Gamma_{g=0}^{[N]}[\cG]
+\frac{\hbar}{4!}\int_{0}^g dg'\tr L:
\left(\fdd{\Gamma_{g'}^{[N]}[\cG]}{\cG}{\cG}+\alpha_{g'}\right)^{-1I}
+\cG:\left(\frac{g}{4!}L-\hf\alpha_g\right):\cG.
\ee
We choose $\alpha=gaL/12$ with a $g$-independent constant $a$ in the
Hartree-term which guides the evolution and seek the effective action in
the form
\be
\Gamma_g[\cG]=\Omega_g[\cG]+\frac{g(1-a)}{4!}\cG:L:\cG.
\ee
The iteration of the functional $\Omega_g[\cG]$,
\be\label{itom}
\Omega^{[N+1]}_g[\cG]=\Omega^{[N]}_{g=0}[\cG]+B[\Omega^{[N]}_g],
\ee
is realized by
\be
B[\Omega_g]=\frac{\hbar}{4!}\int_{0}^g dg'\tr L:
\left[\fdd{\Omega_{g'}[\cG]}{\cG}{\cG}+\frac{g'a}{12}L\right]^{-1I}.
\ee
The starting point is the $g$-independent effective action
$\Omega^{[0]}_{g}=\Gamma^{free}$ and its first iteration,
\be
B[\Omega^{[0]}_g]=\frac{\hbar}{2}
\tr\left\{\left[\ln\left(\tGiI:L^{-1}+\frac{ga}{12}\right)
-\ln(\tGiI:L^{-1})\right]:I\right\},
\ee
yields
\be
\Omega_g^{[1]}[\cG]=\Gamma^{free}[\cG]
+\frac{\hbar}{2}\tr\left[I:\ln\left(\openone+\frac{ga}{12}L:\tG\right)\right].
\ee
The expansion of the logarithm on the right hand side gives
\bea\label{elsoket}
\Gamma_g^{[1]}[\cG]&=&\Gamma^{free}[\cG]-\frac{\hbar}{2}\sum_{n=1}^\infty
\frac1n\left(\frac{-ga}{12}\right)^n\tr[(L:\tG)^{n}]
+\frac{g(1-a)}{4!}\cG:L:\cG,
\eea
since $I:L=L$. This effective action corresponds to the sum
of ring diagrams where the two-particle propagators are joined
by the vertices $g$.

For the choice $a=1$ the right hand side of Eq. \eq{ketiter} is
$\ord{\hbar}$ and each iteration resums a successive order of the
loop expansion to the effective action, c.f. Eq. \eq{sajatener}.
This resummation explains the infinite series in $\Gamma_g^{[1]}[\cG]$.
When $a=0$ is chosen then the right hand side of Eq. \eq{ketiter} is
$\ord{g}$ and each iteration resums an order of the perturbation
expansion in $g$ only. For any other values of $a$ an effective
interaction strength is used in the partial resummation of the
loop expansion (second terms on the right hand sides of Eqs.
\eq{ketiter} and \eq{elsoket}) which is corrected order-by-order
in the perturbation expansion (third terms on the right hand sides of Eqs.
\eq{ketiter} and \eq{elsoket}).

\section{Relations between functional equations}\label{efthe}
We comment in this Section first the relation of the evolution equations with
well-known, non-perturbative iterative methods and next the similarity
of the derivation leading to the evolution equations for the 1PI and 2PI
Green functions, Eqs. \eq{evegyfeleri} and \eq{nintpar}, respectively.

The evolution equations, Eqs. \eq{phi4eveq} and \eq{evolk}
represent exact relations for the generating functionals
which, assuming the proper initial conditions, lead to the
determination of the 1PI and 2PI Green functions, respectively.
It is natural to raise the question about the relation of
these equations to the Schwinger-Dyson functional equation
which is another exact equation satisfied by the Green functions.
Both share a common mathematical origin, they express the
way the path integral for the transition amplitudes
transforms during the change of certain variables \cite{pozsony}.
The Schwinger-Dyson equation, overviewed briefly in Appendix
\ref{schwd}, is a hierarchical set of equations as the
evolution equations. The Schwinger-Dyson equation for the
two and four point irreducible functions, Eq. \eq{sdiprop}
and \eq{41pi}, could be converted into an evolution
equation by acting on them by the operator $\partial_\lambda$. 
But the difference between the resulting equations and Eqs.
\eq{phi4eveq} and \eq{evolk} is that the formers contain
the derivative of the Green functions with respect to the
control parameter in an implicit manner. The evolution equation
scheme is organized in such a manner that the derivatives
are all explicitly expressed in terms of the Green functions
themselves.

There are a few advantages when the evolution equation method is used.
One problem with the Schwinger-Dyson equations is that
it is not clear if its solution is unique. Their iteration
may have different basin of attraction and the identification
of the physical solution might be rather difficult.
As long as the initial conditions are properly given in a
nonphysical, but perturbative domain, the evolution equation
method produces a unique solution.

Let us now compare the evolution equations for the 1PI and 2PI
functions more carefully. A remarkable property of the evolution
equation \eq{phi4eveq} is that it does not lead out from the
space of quadratic functionals. Due to this reason
it is essential to keep the interaction term in the initial condition.
Another way to characterize this feature is to notice that the
structure of the evolution equation reflects the choice of the
$\mu^2$-dependent part  of the bare action only.
The information about the bare interaction is
contained in the initial conditions to the evolution equation.
The 2PI evolution equation, Eq. \eq{evolk}, on the contrary
reflects the $\ord{\varphi^4}$ contact interactions
only and is independent of the $\ord{\varphi^2}$
part of the bare action. In general, the evolution equation for the
$n$-linear expression of the local field variable $\phi_x$ contains
the $\ord{\phi^{n+1}}$ and higher order pieces of the bare action
whose terms up to $\ord{\phi^n}$ appear in the initial conditions only.
An important consequence of this structure is the UV finiteness
of Eq. \eq{evolk}. In fact, the integrand of the formal
one-loop integral on the right hand side is given in terms of the
independent variable $\cG$ and as it stands, Eq. \eq{evolk} is a formal
expression and recovers the well-known divergences for the choice
$\cG_p=\ord{p^{-2}}$ only. The counterterms to the 2PI effective action
depend on the choice of the independent variable $\cG$. It was due to such
a complicated renormalization that we left the discussion of the 2PI
evolution equation on the bare level in Section \ref{seveq}
and counterterms like in Section \ref{fucasa} were not introduced.
As of the initial conditions are
concerned, they must fix that part of the effective action which
is not represented in the evolution equation, in particular
the bare interactions or the free part in  the 1PI
and 2PI cases, respectively.

Finally, the similarity between the 1PI and 2PI evolution equations
represents an opportunity to find an alternative to the matching
strategy for the introduction of effective theories.
This similarity suggests to replace the composite operator
$\varphi_x\varphi_y$ by a bilocal effective field $\psi_{x,y}$ and
inquire about the corresponding effective theory given by the
action $S_\mr{eff}[\psi]$.
The construction of an effective theory for local composite operators
is well established by means of the perturbation expansion \cite{compeff},
or of non-perturbative scheme \cite{nonpeff}. Our starting point
for the derivation of the effective theory for the bilocal field $\psi_{x,y}$
is the generating functional $\tW[J]$ of Eq. \eq{www} for the connected Green
functions of the composite operator $\varphi_x\varphi_y$.
The free bare effective action, $S^{(0)}_\mr{eff}[\psi]$, is chosen in such
a manner that it reproduces the free propagation of two identical particles
controlled by $\psi_{x,y}$,
\be\label{wwwpsi}
\int\cd{\varphi}e^{-\frac1{\hbar}\int_x[-\hf\varphi_x\Box_x\varphi_x
+\hf m^2\varphi^2_x]
+\frac1{2\hbar}\varphi\cdot J\cdot\varphi}
=\int\cd{\psi}e^{-\frac1{\hbar}S^{(0)}_\mr{eff}[\psi]+\frac1{2\hbar}\psi:J}.
\ee
The interactions can be incorporated by adding the piece
$g\int_x\psi_{x,x}^2/4!$ to the effective action. The equivalence of the
interactive theories, expressed by the relation
\be\label{wwwpsii}
e^{\frac1{\hbar}\tW[J]}=\int\cd{\psi}e^{-\frac1{\hbar}S^{(0)}_\mr{eff}[\psi]
-\frac{g}{4!\hbar}\int_x\psi_{x,x}^2+\frac1{2\hbar}\psi:J},
\ee
can be verified by acting with the differential operator
\be
e^{-\frac{\hbar g}{4!}\int_x\fdd{}{\hf J_{x,x}}{\hf J_{x,x}}}
\ee
on both sides of Eq. \eq{wwwpsi}.

The first step of this argument, the construction of the free action
$S^{(0)}_\mr{eff}[\psi]$ is far more complicated than the second step,
the inclusion of the interactions. This is not by accident,
the non-interacting dynamics of equivalent particles has to include
the symmetrization/antisymmetrization of the quantum states. This
step introduces long range, topological correlations among the
otherwise non-interacting particles. These correlations represent
a far more difficult problem than the local interactions.
This difficulty is particularly explicit in the construction
of the density functional \cite{dft} for non-interacting particles
when the effective action has strongly non-local vertices of
infinitely high order  without natural small parameter to justify any
truncation. When the non-interacting particles are fermions
this procedure amounts to bosonisation.

The difficulty of obtaining the free effective action is
avoided when one uses the evolution
equation method and pays the price of having an effective action
for a bilocal variable $\cG_{x,y}$ rather than the density functional
which depends on the local, diagonal part $\cG_{x,x}$ only.
The complications mentioned are then hidden in the logarithmic
function of the initial condition Eq. \eq{initk}.

The generalization of this construction
for higher order composite operators is possible, as well.
The only condition is that the interaction Lagrangian for the
elementary fields must be expressible in terms of the
composite fields. The difficult question,
the issue of the choice of the interactions in the effective
theory can be hidden in the initial conditions for the
effective action for the composite fields. This latter
can partially be taken over from the elementary theory without
interactions (aspects of the statistics) and partially be
generated by the evolution equation (aspects of the
interactions) which provides an iterative improvement of the
effective theory (cf. Section \ref{itersol}).

\section{Summary}\label{su}
The functional renormalization group strategy was studied in this paper in the
internal space in the framework of the scalar $\phi^4$ model
where either the mass or the coupling strength can be evolved.
The mass-controlled suppression of the amplitude
of the fluctuations realizes the functional generalization of the
Callan-Symanzik scheme for the bare Green functions. The
appropriate split of the evolution equation into the sum of two
pieces provides a non-perturbative BPHZ-like renormalization
of the 1PI Green functions.

The control of the amplitude of the fluctuations by
the interactions leads to the evolution equation generated by the
change of the coupling strength. This scheme was exploited in order to
obtain the evolution equation for the 2PI effective action.

Finally, the similarity of the evolution equations in these schemes
was used to construct an effective bilocal theory for the connected
propagator.

The advantage of the method, compared to the traditional
external space renormalization group schemes, lies in its flexibility
in choosing the parameter to evolve. This
feature allows one to  trace the dependence on certain scales in a model
possessing several scales or on any of the coupling constants. This latter
is important when symmetries prevent us to use momentum-space
cutoff.

\appendix
\section{Notation, conventions}\label{not}
The space-time and Fourier-space integrals are defined as
\bea
\int_x=\int d^dx=a^d\sum_x,&~~~~&
\int_p=\int\frac{d^dp}{(2\pi)^d}
=\sum_p,~~~~
f_x=\int_pf_pe^{ipx},
\eea
and the Fourier transform of a translation invariant propagator
$G_{x,y}$ is defined as $\delta_{p+q,0}G_p=G_{p,q}$.
The composite index $X=(x,y)$, proves to be useful, e.g.
$\int_{x,y}A_{x,y}B_{x,y}=\int_XA_XB_X$. The integration is
frequently shown as scalar product, $f\cdot g=\int_xf_xg_x$,
$A:B=\int_{x,y}F_{x,y}G_{x,y}$.
The repeated indices are automatically summed/integrated over unless
written explicitly otherwise. 

The functional derivative is defined in $d$-dimensional lattice as
\be
\fd{}{\phi_x}=\frac{1}{a^d}\frac{\partial}{\partial\phi^L_x}
\ee
where $\phi^L$ is the lattice field variable where the factor $a^{-d}$ is needed
in order to satisfy the relation
\be
\fd{}{\phi_y}\int_xf(\phi_x)=f'(\phi_y).
\ee
In an analogous manner we have
\be
\fd{}{\phi_p}=V\frac{\partial}{\partial\phi^L_p}.
\ee
Our convention for the derivative with respect to a symmetrical matrix
$\cG^\mr{sym}$ is
\be\label{gfder}
\fd{}{\cG^\mr{sym}_{x,y}}=\begin{cases}
\frac{1}{2a^{2d}}\left(
\frac{\partial}{\partial\cG^{\mr{non-sym}L}_{x,y}}
+\frac{\partial}{\partial\cG^{\mr{non-sym}L}_{y,x}}\right)&x\not=y,\cr
\frac{1}{a^{2d}}\frac{\partial}{\partial\cG^{\mr{sym}L}_{x,y}}&x=y,
\end{cases}
\ee
where $\cG^\mr{non-sym}$ is a non-symmetrical matrix. It
allows us the use of the chain rule
\be
\fd{}{A_{x,y}}=\int_{u,v}\fd{B_{u,v}}{A_{x,y}}\fd{}{B_{u,v}}.
\ee

\section{One-loop effective action}\label{olefac}
The evolution equation for the 2PI effective action requires the knowledge
of initial conditions imposed at vanishing or infinitesimal coupling
strength. The loop-expansion for path integral Eq. \eq{www} has two
non-vanishing orders for $g=0$, therefore these orders will be retained below.

\subsection{$\ord{\hbar^0}$}
The generating functionals are given by
\bea
\tW^{(0)}[J]&=&-S[\phi_0[J]]+\hf\phi_0[J]\cdot J\cdot\phi_0[J],\nn
\tGa^{(0)}[\cG]&=&S[\phi_0[J]]-\hf\phi_0[J]\cdot J\cdot\phi_0[J]
+\hf\tr\cG^t\cdot J,
\eea
in the tree approximation where the stationary point $\phi_0[J]$ satisfies
the equation
\be
\fd{S[\phi]}{\phi}_{|\phi=\phi_0[J]}=J\cdot\phi_0[J],
\ee
and the source $J$ is related to the Green function $\cG$ via
\be
\cG_{x,y}=\phi_{0~x}[J]\phi_{0~y}[J],
\ee
i.e.
\be
\phi_{0~x}[J]=\sqrt{\cG_{x,x}}.
\ee
Notice that $\phi_0[J]$ is non-vanishing and is given by this equation
in the phase with broken symmetry  only. The elimination of the source
$J$ yields
\bea
\tGa^{tree}[\cG]&=&S[\phi_0[J]]
=\int_x\left[-\frac18(\ln\cG_{x,x})\Box_x\cG_{x,x}
+\frac{\mu^2}{2}\cG_{x,x}+\frac{g}{4!}(\cG_{x,x})^2\right].
\eea
In the symmetrical phase without tree-level condensate $\tGa^{tree}[\cG]=0$
holds.

\subsection{$\ord{\hbar}$}
We write $\phi=\phi_0+\eta$, expand the exponent of
the integrand in powers of $\eta$ keeping the terms up to
$\ord{\eta^2}$ and find
\be
\tW^{(1)}[J]=-S[\phi_0[J]]+\hf\phi_0[J]\cdot J\cdot\phi_0[J]
-\hf\hbar\tr\ln\left[\fdd{S[\phi]}{\phi}{\phi}_{|\phi_0[J]}
- J\right].
\ee
We define the connected two-point function $G$ by
\bea
\cG_{x,y}&=&\phi_{0~x}[J]\phi_{0~y}[J]+\hbar G_{x,y}
=\fd{W^{(1)}[J]}{\hf J_{x,y}}
=\phi_{0~x}[J]\phi_{0~y}[J]+\hbar\left[\fdd{S[\phi]}{\phi}{\phi}
_{|\phi_0[J]}-J\right]^{-1}_{y,x}
\eea
which gives
\be
G_{x,y}=\left[\fdd{S[\phi]}{\phi}{\phi}_{|\phi_0[J]}- J
\right]^{-1}_{y,x},
\ee
and
\be\label{olga}
\tilde\Gamma^{(1)}[\cG]=\Gamma^{tree}[\cG]
+\frac{\hbar}{2}\tr\ln G^{-1}+\frac{\hbar}{2}\tr\left[
G^{t}\cdot\fdd{S[\phi]}{\phi}{\phi}\right]
\ee
up to a $\cG$-independent constant, c.f. Ref. \cite{cjt}.
In order to avoid the problem of non-convexity, we assume
that the vacuum has no condensate, $\phi_0[J]=0$, i.e. $\cG=\hbar G$.

\section{Inverse of symmetric 4-point functions}\label{fourier}
The quadratic part of the free effective action is
\be
\tGiI_{(x,y),(u,v)}=\frac{1}{4}
\int_{p,q}\cG_p^{-1}\cG_q^{-1}\left(e^{ip(y-u)+iq(v-x)}
+e^{ip(y-v)+iq(u-x)}\right),
\ee
for the translation invariant case.
Therefore we seek the inversion of the operators of the form
\be\label{reginv}
A_{(x,y),(u,v)}=\hf\int_{p,q}A_{p,q}
\left(e^{ip(x-u)+iq(y-v)} + e^{ip(x-v)+iq(y-u)}\right),
\ee
with symmetries $A_{(x,y),(u,v)}= A_{(u,v),(x,y)}=A_{(x,y),(v,u)}$ which
require $A_{p,q}= A_{-q,-p}=A_{q,p}$. It is easy to see that $A^{-1}$
cannot be written in the form \eq{reginv}. But such operators
can be inverted when they are acting on the space of symmetric matrices.
In fact, the relation
\be
\hf(\delta_{x,u}\delta_{y,v}+\delta_{x,v}\delta_{y,u})
=\hf\int_{p,q}A_{p,q}A^{-1}_{p,q}\left(e^{ip(y-v)+iq(u-x)}
+e^{ip(y-u)+iq(v-x)}\right)
\ee
yields $I_{p,q}=1$, $A^{-1I}_{p,q}=1/A_{p,q}$ and
\be
\tGiI_{p,q}=\hf\cG^{-1}_p\cG_q^{-1},~~~~\tG_{p,q}=2\cG_p\cG_q
\ee
for the free two-particle propagator.

\section{Equation of motion}\label{schwd}
The Schwinger-Dyson equation is obtained in this Appendix for the two- and
four-point 1PI functions. The absence of the condensate is assumed, in order
to simplify the expressions.

\subsection{Elementary field variable}\label{efv}
The path integral
\be
Z[j]\equiv e^{W[j]}=\int\cd{\varphi}e^{-S[\varphi]+\varphi\cdot j}
\ee
yields the Schwinger-Dyson equation
\be
\left[\fd{S[\phi]}{\phi}_{|\phi=\fd{}{j}}-j\right]e^{W[j]}=0,
\ee
which can be written as
\be\label{sdf}
\left[\fd{S[\phi]}{\phi}_{|\phi=\fd{}{j}+\fd{W[j]}{j}}-j\right]
\cdot\openone=0
\ee
by means of the identity
\be
\left[\fd{}{j},e^{W[j]}\right]=\fd{W[j]}{j}e^{W[j]} .
\ee
For
\be
S=\hf\varphi\cdot G_0^{-1}\cdot\varphi+\frac{g}{4!}\int_x\varphi^4
\ee
Eq. \eq{sdf} yields the Schwinger-Dyson equation
\be\label{sd}
0=\phi_x+\frac{g}{6}\int_y G_{0~x,y}\left[
\phi_y^3+3\phi_y\left(\fdd{\Gamma[\phi]}{\phi}{\phi}\right)^{-1}_{y,y}
+\fddd{W[j]}{j_y}{j_y}{j_y}\right]-(G_0\cdot j)_x .
\ee
The first functional derivative of this equation is
\be
\fdd{W[0]}{j_x}{j_y}=G_{0~x,y}
-\frac{g}{2}\int_zG_{0~x,z}\fdd{W[0]}{j_z}{j_y}\fdd{W[0]}{j_z}{j_z}
-\frac{g}{6}\int_zG_{0~x,z}\frac{\delta^4W[0]}
{\delta j_z\delta j_z\delta j_z\delta j_y}
\ee
for $j=0$. We shall express this relation
by means of the 1PI vertex functions. By multiplying with $G_0^{-1}$ and
the inverse propagator we first arrive at
\be\label{propeq}
\fdd{\Gamma[0]}{\phi_x}{\phi_y}=G^{-1}_{0~x,y}
+\frac{g}{2}\delta_{x,y}\fdd{W[0]}{j_x}{j_x}
+\frac{g}{6}\int_z\frac{\delta^4W[0]}
{\delta j_x\delta j_x\delta j_x\delta j_z}\fdd{\Gamma[0]}{\phi_z}{\phi_y}.
\ee
In order to express the last term on the right hand side in terms of the
effective action, one acts with the operator
\be
\fd{}{j}=\fd{\phi}{j}\cdot\fd{}{\phi}=\fdd{W[j]}{j}{j}\cdot\fd{}{\phi}
=\left(\fdd{\Gamma}{\phi}{\phi}\right)^{-1}\cdot\fd{}{\phi}
\ee
on the identity
\be
\openone=\fdd{W[j]}{j}{j}\cdot\fdd{\Gamma[\phi]}{\phi}{\phi}
\ee
and finds
\bea
0&=&\int_z
\fddd{W[j]}{j_{x_1}}{j_{x_2}}{j_z}\fdd{\Gamma[\phi]}{\phi_z}{\phi_{x_3}}
+\int_{\{z\}}\fdd{W[j]}{j_{x_1}}{j_{z_1}}
\fddd{\Gamma[\phi]}{\phi_{z_1}}{\phi_{z_2}}{\phi_{x_3}}
\fdd{W[j]}{j_{z_2}}{j_{x_2}},
\eea
where the shorthand notation $\int_{\{z\}}=\int_{z_1,\ldots,z_n}$ 
has been introduced. Multiplication by the second functional derivative
of $W[j]$ in the index $x_3$ gives
\be\label{haromder}
\fddd{W[j]}{j_{x_1}}{j_{x_2}}{j_{x_3}}
=-\int_{\{z\}}\fdd{W[j]}{j_{x_1}}{j_{z_1}}\fdd{W[j]}{j_{x_2}}{j_{z_2}}
\fdd{W[j]}{j_{x_3}}{j_{z_3}}
\fddd{\Gamma[\phi]}{\phi_{z_1}}{\phi_{z_2}}{\phi_{z_3}}.
\ee
A further derivation leads to
\bea
\frac{\delta^4W[j]}{\delta j_{x_1}\delta j_{x_2}\delta j_{x_3}\delta j_{x_4}}
&=&\int_{\{z\}}
\biggl[-\fdd{W[j]}{j_{x_1}}{j_{z_1}}\fdd{W[j]}{j_{x_2}}{j_{z_2}}
\fdd{W[j]}{j_{x_3}}{j_{z_3}}\fdd{W[j]}{j_{x_4}}{j_{z_4}}
\frac{\delta^4\Gamma[\phi]}{\delta\phi_{z_1}\delta\phi_{z_2}
\delta\phi_{z_3}\delta\phi_{z_4}}\nn
&&+\fdd{W[j]}{j_{x_1}}{j_{z_4}}\fdd{W[j]}{j_{x_4}}{j_{z_5}}
\fdd{W[j]}{j_{z_1}}{j_{z_6}}
\fddd{\Gamma[\phi]}{\phi_{z_4}}{\phi_{z_5}}{\phi_{z_6}}
\fdd{W[j]}{j_{x_2}}{j_{z_2}}\fdd{W[j]}{j_{x_3}}{j_{z_3}}
\fddd{\Gamma[\phi]}{\phi_{z_1}}{\phi_{z_2}}{\phi_{z_3}}\nn
&&+(x_1\longleftrightarrow x_3)+(x_1\longleftrightarrow x_4)\biggr]
\eea
which finally allows us to write Eq. \eq{propeq} as
\bea\label{sdiprop}
\fdd{\Gamma[0]}{\phi_x}{\phi_y}&=&G^{-1}_{0~x,y}+\frac{g}{2}\delta_{x,y}
\left(\fdd{\Gamma[0]}{\phi}{\phi}\right)^{-1}_{x,x}\nn
&&-\frac{g}{6}\int_{\{z\}}
\left(\fdd{\Gamma[0]}{\phi}{\phi}\right)^{-1}_{x,z_1}
\left(\fdd{\Gamma[0]}{\phi}{\phi}\right)^{-1}_{x,z_2}
\left(\fdd{\Gamma[0]}{\phi}{\phi}\right)^{-1}_{x,z_3}
\frac{\delta^4\Gamma[0]}{\delta\phi_{z_1}\delta\phi_{z_2}
\delta\phi_{z_3}\delta\phi_y}\nn
&&+\frac{g}{2}\int_{\{z\}}\left(\fdd{\Gamma[0]}{\phi}{\phi}\right)^{-1}_{x,z_4}
\left(\fdd{\Gamma[0]}{\phi}{\phi}\right)^{-1}_{z_1,z_6}
\fddd{\Gamma[0]}{\phi_{z_4}}{\phi_y}{\phi_{z_6}}
\left(\fdd{\Gamma[0]}{\phi}{\phi}\right)^{-1}_{x,z_2}
\left(\fdd{\Gamma[0]}{\phi}{\phi}\right)^{-1}_{x,z_3}
\fddd{\Gamma[0]}{\phi_{z_1}}{\phi_{z_2}}{\phi_{z_3}}.
\eea

\subsection{Product of two field variables}
The path integral
\be
\tZ[J]\equiv e^{\tW[J]}=\int\cd{\varphi}e^{-S[\varphi]
+\hf\varphi\cdot J\cdot \varphi}
\ee
leads to the relations
\bea
\fd{\tW[0]}{\hf J_{1,2}}&=&\fdd{W[0]}{j_1}{j_2}=G^c_{1,2}=\cG_{\mr{v}~1,2},\nn
\fdd{\tW[0]}{\hf J_{1,2}}{\hf J_{3,4}}&=&K_{(1,2),(3,4)},\nn
\fddd{\tW[0]}{\hf J_{1,2}}{\hf J_{3,4}}{\hf J_{5,6}}
&=&\frac{\fddd{\tZ[0]}{\hf J_{1,2}}{\hf J_{3,4}}{\hf J_{5,6}}}{\tZ[0]}
-K_{(1,2),(3,4)}\cG_{\mr{v}~5,6}-K_{(1,2),(5,6)}\cG_{\mr{v}~3,4}
-K_{(3,4),(5,6)}\cG_{\mr{v}~1,2}\nn
&&-\cG_{\mr{v}~1,2}\cG_{\mr{v}~3,4}\cG_{\mr{v}~5,6},
\eea
where the two-particle propagator
\be
K=\left(\fdd{\tGa[\cGv]}{\cG}{\cG}\right)^{-1}
\ee
has been introduced. (For the sake of simplicity the space-time
 coordinates $x_n$ are replaced
by the corresponding indices $n$.)
The analogue of Eq. \eq{haromder} is
\be
\fddd{\tW[J]}{\hf J_{X_1}}{\hf J_{X_2}}{\hf J_{X_3}}
=-\left(\fdd{\tGa[\cG]}{\cG}{\cG}\right)^{-1}_{X_1,Z_1}
\left(\fdd{\tGa[\cG]}{\cG}{\cG}\right)^{-1}_{X_2,Z_2}
\left(\fdd{\tGa[\cG]}{\cG}{\cG}\right)^{-1}_{X_3,Z_3}
\fddd{\tGa[\cG]}{\cG_{Z_1}}{\cG_{Z_2}}{\cG_{Z_3}}.
\ee
Therefore, we have
\bea
G^c_{1,2}&=&\cG_{\mr{v}~1,2},\nn
G^c_{1,2,3,4}&=&K_{(1,2),(3,4)}-\tG_{\mr{v}~(1,2),(3,4)}\nn
G^c_{1,2,3,4,5,6}&=&-K_{(1,2),Z_1}K_{(3,4),Z_2}K_{(5,6),Z_3}
\fddd{\Gamma[\cGv]}{\cG_{Z_1}}{\cG_{Z_2}}{\cG_{Z_3}}\nn
&&-[K_{(1,2),(3,5)}G^c_{4,6}+K_{(1,2),(4,5)}G^c_{3,6}
+K_{(1,2),(3,6)}G^c_{4,5} +K_{(1,2),(4,6)}G^c_{3,5}
+G^c_{1,3,4,5}G^c_{2,6}+G^c_{2,3,4,5}G^c_{1,6}\nn
&&+G^c_{1,3,4,6}G^c_{2,5}+G^c_{2,3,4,6}G^c_{1,5}
+G^c_{1,3,5,6}G^c_{2,4}+G^c_{2,3,5,6}G^c_{1,4}
+G^c_{1,4,5,6}G^c_{2,3}+G^c_{2,4,5,6}G^c_{1,3}]
\eea
with $\tG_{\mr{v}~X,Y}$ the free two-particle propagator.
The first functional derivative of the Schwinger-Dyson equation
can be written as
\be\label{21pi}
\Sigma_{x,y}=\frac{g}{2}\delta_{x,y}\cG_{\mr{v}~x,x}
+\frac{g}{6}\int_z(K_{(x,x),(x,z)}-\tG_{\mr{v}~(x,x),(x,z)})\cGi_{\mr{v}~z,y}.
\ee
Notice that the leading order perturbative result is recovered for
$K=\tG_{\mr{v}}$ only. When the two-particle sector is non-trivial by the
choice of the independent variable $\cG$ then the one-particle
sector of the theory is naturally modified.

In a similar manner the third functional derivative of the Schwinger-Dyson
equation reads as
\bea\label{41pi}
K_{(x,y),(u,v)}&=&\tG_{\mr{v}~(x,y),(u,v)}+\frac{g}{6}\int_zG_{0~x,z}\biggl(
K_{(z,z),Z_1}K_{(z,y),Z_2}K_{(u,v),Z_3}
\fddd{\Gamma[\cGv]}{\cG_{Z_1}}{\cG_{Z_2}}{\cG_{Z_3}}\nn
&&+K_{(z,z),(z,u)}\cG_{\mr{v}~y,v}+K_{(z,z),(z,v)}\cG_{\mr{v}~y,u}
-K_{(z,z),(u,v)}\cG_{\mr{v}~z,y}-K_{(z,y),(u,v)}\cG_{\mr{v}~z,z}
+\cG_{\mr{v}~z,z}\tG_{\mr{v}~(z,y),(u,v)}\biggr).\nn
\eea


\begin{thebibliography}{99}
\bibitem{trrg} K.G. Wilson, J. Kogut, \Journal{\PREPC}{12}{77}{1974};
K.G. Wilson, \Journal{\RMP}{47}{773}{1975}; ibid.
 \Journal{\RMP}{55}{583}{1983}.
\bibitem{wh} F.J.Wegner, A.Houghton, \Journal{\PRA}{8}{401}{1973}.
\bibitem{polch} J. Polchinski, \Journal{\NPB}{231}{269}{1984}.
\bibitem{morrp} T. Morris, {\em Int. J. Mod. Phys.} A {\bf 9}, 2411 (1994).
\bibitem{wett} C. Wetterich, \Journal{\PLB}{301}{90}{1993}.
\bibitem{ca} C. G. Callan, \Journal{\PRD}{2}{1541}{1970}.
\bibitem{sy} K. Symanzik, \Journal{\CMP}{18}{227}{1970}.
\bibitem{int} J. Alexandre, J. Polonyi, \Journal{\AP}{288}{37}{2001}.
\bibitem{sim} M. Simionato, \Journal{\IJMP}{A15}{2121}{2000}.
\bibitem{bp} N.N. Bogoljubov, O. Parasiuk, \Journal{\AM}{97}{227}{1957}.
\bibitem{hepp} K. Hepp, \Journal{\CMP}{2}{301}{1966}.
\bibitem{zimm} W. Zimmermann, \Journal{\CMP}{15}{208}{1969}.
\bibitem{cjt} J. M. Cornwall, R. Jackiw, E. Tomboulis,
{\em Phys. Rev.} D {\bf 10}, 2428 (1974).
\bibitem{hay} R.W. Haymaker, {\em Riv. Nuovo Cim.} {\bf 14}, 1 (1991).
\bibitem{qed} J. Alexandre, J. Polonyi, K. Sailer, 
\Journal{\PLB}{531}{316}{2002}.
\bibitem{dft} J. Polonyi, K. Sailer, \Journal{\PRB}{66}{155113}{2002}.
\bibitem{loc} J. Polonyi, \Journal{\PRB}{66}{014202}{2003}.
\bibitem{hel} H. Hellmann, {\em Einf\"uhrung in die Quantenchemie}
(Deuticke, Leipzig, 1937).
\bibitem{fey} R.P. Feynman, {\em Phys. Rev.} {\bf 56}, 340 (1939).
\bibitem{optsb} S. B. Liao, J. Polonyi, M. Strickland,
\Journal{\NPB}{567}{493}{2000}.
\bibitem{optdl} D. F. Litim, \Journal{\PLB}{486}{92}{2000}.
\bibitem{pozsony} J. Polonyi, {\em Central European J. of Phys.}
{\bf 1}, 1 (2003).
\bibitem{compeff} J. Collins, {\em Renormalization} 
(Cambridge Univ. Press, Cambridge,1984).
\bibitem{nonpeff} J. Polonyi, K. Sailer,{\em Phys. Rev.} D {\bf 63}, 105006 (2001).
\end{thebibliography}
\end{document}